\documentclass[12pt,a4paper]{article}
  \usepackage{a4wide}
  \usepackage{latexsym}
  \usepackage{epsf}
  \usepackage{amssymb}
  \usepackage{graphicx}
  \usepackage{amsmath, cite}
  \usepackage{amsmath,amssymb,amsthm}
  \usepackage{verbatim}
  \usepackage{hyperref}
  \usepackage{color}
  \usepackage{mathtools}
  \usepackage{soul,xcolor}

\DeclarePairedDelimiter\abs{\lvert}{\rvert}%
\usepackage{makecell}

\newcommand{\be}{\begin{equation}}
\newcommand{\ee}{\end{equation}}

\newenvironment{aleq}
    {\begin{equation}\begin{aligned}}
    {\end{aligned}\end{equation}\ignorespacesafterend}

\begin{document}
\numberwithin{equation}{section}

\vspace{0.4cm}
\begin{center}

{\LARGE \bf{Aspects of AdS flux vacua with integer conformal dimensions}}\\

\vspace{2 cm} {\large Fien Apers}\\
\vspace{1 cm} {\small\slshape Rudolf Peierls Centre for Theoretical Physics,
Beecroft Building, \\Clarendon Laboratory, Parks Road, University of Oxford, OX1 3PU, UK}\\
\vspace{0.5 cm} {\small\slshape fien.apers@physics.ox.ac.uk}\\


\vspace{1cm}

{\bf Abstract} \end{center} {The DGKT vacua are a class of AdS$_4$ flux vacua showing full moduli stabilization, parametric control, and a parametric separation of scales. The particular masses of the moduli remarkably give rise to integer conformal dimensions in the light spectrum of the would-be holographic duals. In this note, we comment on two  properties for AdS flux vacua with integer conformal dimensions. First, there are polynomial spacetime-dependent shift symmetries for the moduli. Secondly, the leading scalings of the central charge and the moduli can be directly deduced from the near-horizon geometry of stacks of orthogonally-intersecting D-brane domain walls dual to the unbounded fluxes. We illustrate this in a couple of examples of AdS$_4$ and AdS$_3$ parametric flux vacua.}

\newpage

\tableofcontents
\section{Introduction}
Realistic string theory vacua should have full moduli stabilization and a separation of scales between the Hubble scale $L_H$ and the Kaluza-Klein length scale $L_{KK}$, corresponding to the diameter of the internal manifold,
\begin{aleq}
    \dfrac{L_{KK}}{L_H} \ll 1.
\end{aleq}
The DGKT vacua \cite{DeWolfe:2005uu, Camara:2005dc} are a class of $\mathcal{N}=1$ AdS$_4$ flux vacua with full moduli stabilization, obtained from compactifying massive type IIA string theory on a Calabi-Yau threefold with fluxes and intersecting O6-planes. As there is unbounded flux $\abs{F_4}\sim N$, parametric control
\begin{aleq}
    g_s \sim N^{-3/4}, \quad vol_S \sim N^{3/2},
\end{aleq}
with $g_s$ the string coupling and $vol_S$ the internal volume in string frame,
and parametric scale separation 
\begin{aleq}\label{SS}
   \dfrac{L_{KK}}{L_H} \sim N^{-1/2}
\end{aleq}
can be achieved by sending this flux to infinity. 

There is however no full 10-dimensional understanding of the backreaction of the O6-planes on the internal geometry (for recent progress on this, see e.g. \cite{Cribiori:2021djm,Junghans:2020acz, Marchesano:2020qvg}), and scale-separated supersymmetric AdS vacua are conjectured to be in the swampland \cite{Lust:2019zwm}. Studying the holographic CFT duals is a promising alternative approach for checking the consistency of these AdS vacua \cite{Conlon:2018vov, Conlon:2020wmc, Lust:2022lfc}.

The would-be CFT duals of the DGKT vacua do have some very peculiar properties. First, scale-separated AdS vacua in general should be dual to CFTs with a large gap in the spectrum of single-trace primaries, and such CFTs are unknown \cite{Collins:2022nux}.
Also, the central charge should scale like
\begin{aleq}
    c_{4d} \sim N^{9/2},
\end{aleq}
which is a higher scaling than in any known holographic set-up\cite{Aharony:2008wz, Banks:2006hg}, which goes up to $N^3$ \cite{Maldacena:1997re, Aharony:1999ti}. 
Moreover, the spectrum of light operators, dual to the moduli, would consist fully of integer conformal dimensions \cite{Conlon:2021cjk} (see also \cite{Ning:2022zqx, Plauschinn:2022ztd, Quirant:2022fpn, Quirant:2022ybh}). The K\"ahler moduli and dilaton are dual to $h^{1,1}_{-}$ operators with $\Delta = 6$ and one operator with $\Delta = 10$.
These integers are universal \cite{Marchesano:2020uqz, Apers:2022tfm}: independent of the values of the fluxes and microscopic details of the Calabi-Yau internal manifold. As there is no extended supersymmetry in these CFTs (3d $\mathcal{N}=1$), the presence of integers is very surprising and requires an explanation.

In this work we argue that integer dimensions are interesting, with two remarkable properties for AdS flux vacua with integers:
\begin{itemize}
    \item Moduli fields in AdS flux vacua, dual to light operators with integer conformal dimensions, do have polynomial spacetime-dependent shift symmetries.
    \item  It is possible to obtain the leading scalings of the central charge, string coupling and internal volume from the near-horizon geometry of a large number $N$ of orthogonally-intersecting D-brane domain walls.
\end{itemize}
We heavily lean on \cite{Kounnas:2007dd} about a domain wall - flux correspondence in AdS$_4$ flux vacua. It is explained there how
the fluxes ($F_4$, $H_3$, $F_0$) in DGKT can be interchanged for domain walls (wrapped D4-, NS5- and D8-branes), interpolating between AdS and flat spacetime, and how a AdS$_4 \times T_6$ geometry can be recovered as the near-horizon limit of these domain walls.
We will just focus on the three stacks of $N$ orthogonally intersecting D4-branes of this set-up, dual to the unbounded flux $F_4 \sim N$, and observe how the large $N$ scalings in DGKT 
\begin{aleq}
    c\sim N^{9/2}, \quad g_s \sim N^{-3/4}, \quad vol_S \sim N^{3/2},
\end{aleq}
can be directly deduced from the near-horizon geometry of these domain walls.
For other parametric AdS$_4$ flux vacua with an integer $\Delta = 6$ in the spectrum, we observe how the large $N$ scalings can be derived in a similar way.

Because the space of 2d CFTs is better understood, it might be easier to bootstrap the holographic duals of AdS$_3$ vacua. Scale-separated AdS$_3$ vacua with minimal supersymmetry were found in \cite{Farakos:2020phe} from compactification of massive IIA string theory on G2 holonomy spaces. The ingredients are very similar as for the DGKT vacua, with unbounded $F_4$-fluxes, and bounded $H_3$- and $F_0$-fluxes in combination with intersecting O6-planes. Now O2-planes are needed as well. Thanks to the unbounded $F_4$-flux, parametric control
\begin{aleq}\label{ftv1}
    g_s\sim N^{-3/4}, \quad vol_S \sim N^{7/4},
\end{aleq}
and parametric scale separation \eqref{SS} are achieved in the $N\rightarrow \infty$ limit. 
The central charge of the CFT$_2$ duals should now be
\begin{aleq}\label{ftv2}
    c_{3d} \sim N^{4},
\end{aleq}
and the light spectrum consists of irrational dimensions \cite{Apers:2022zjx}. In this case, it is not possible to obtain the scalings \eqref{ftv1}, \eqref{ftv2} from the near-horizon geometry of a set of orthogonally intersecting D4-brane domain walls.

There are however other AdS$_3$ vacua in IIB string theory \cite{Emelin:2021gzx} for which the holographic brane set-up might be much simpler and more alike the 4d examples with $\Delta = 6$ we discuss.
The internal manifold is a G2-structure manifold, and $F_7$- and $F_3$-fluxes are used \cite{Emelin:2021gzx}. Parametric scale separation can not be achieved for these vacua, but by taking $F_7\sim N$ to be large, parametric control can be obtained. Interestingly, in a two-moduli truncation, the conformal dimensions are integer and rational ($\Delta_1 = 4$ and $\Delta_2 = 20/7$). Moreover, the leading scalings are now obtainable from the near-horizon limit of a stack of $N$ D1-branes, dual to the $F_7$-flux. So despite not showing a separation of scales, an exploration the holographic duals of these vacua, instead of the ones with irrational dimensions \cite{Farakos:2020phe} mentioned above, might bring us closer to an understanding of the DGKT duals. The properties of the different AdS vacua we discuss are summarized in Table \ref{summary}.
\begin{table}[h!]
        \begin{center}
            \begin{tabular}{ | l | l | l |l |}
            \hline
             & IIA AdS$_4$ \cite{DeWolfe:2005uu} & IIA AdS$_3$ \cite{Farakos:2020phe} & IIB AdS$_3$ \cite{Emelin:2021gzx}  \\
            \hline
            parametric control & yes & yes & yes\\
            \hline
           parametric scale separation & yes & yes & no\\
            \hline
            integer dimension $\Delta = 2d$ & yes & no & yes\\
            \hline
            simple brane set-up & yes & no & yes\\
            \hline
        \end{tabular}
    \caption{Properties of different parametric AdS vacua in $(d+1)$ dimensions.}
    \label{summary}
    \end{center}
\end{table}

The outline is as follows.
In Section 2, the relation between shift symmetries and integer dimensions is explained. Then, in Section 3, we first describe how to derive large $N$ scalings from D-brane domain walls in general in Section \ref{generalflux}. We apply this in Section \ref{4ddomainwalls} to the DGKT vacua, both in massive ($F_0 \neq 0$) and in a double T-dual massless ($F_0=0$) form. We also briefly touch upon more general parametric AdS$_4$ vacua which have a $\Delta = 6$ in the spectrum. In Section \ref{3ddomainwalls} the large $N$ scalings in IIA and IIB AdS$_3$ vacua are discussed, and we conclude in Section 4. 
\section{Polynomial Shift Symmetries}
The conformal dimensions of the light fields in the DGKT CFT duals are \cite{Collins:2022nux, Apers:2022tfm}
\begin{aleq}
\Delta_\phi = ( 6, \dots 6, 10), \quad \Delta_a = (5, \dots 5, 11),
\end{aleq}
for the $h^{1,1}_{-}+1$ saxions and axions respectively, and
\begin{aleq}
\Delta_\phi = ( 2, \dots 2), \quad \Delta_a = (3, \dots 3),
\end{aleq}
for the $h^{2,1}$ complex structure moduli and their axions. All dimensions are integer.

As conformal dimensions in AdS$_{d+1}$/CFT$_d$ are obtained from the masses via a quadratic equation
\begin{aleq}
    \Delta (\Delta - d) = m^2 R_{AdS}^2,
\end{aleq}
it is surprising to get integers and one expects there to be a symmetry at work. Massless scalars, with $\Delta = 3$, for example, enjoy a constant shift symmetry. Could there be different shift symmetries for other integers?

For a free massless field in Minkowski space, there is not only the constant shift symmetry
\begin{aleq}
\phi \rightarrow \phi + c, \quad c \in \mathbb{R},
\end{aleq}
but also an infinite sequence of shift symmetries that are polynomial in the spacetime coordinates $X^\mu$,
\begin{aleq}
\phi \rightarrow \phi + c + c_{\mu}X^\mu + c_{\mu \nu} X^\mu X^\nu + \dots,
\end{aleq}
where $c_{\mu_1 \dots \mu_k}$ are rank-$k$ symmetric traceless constant tensors \cite{Hinterbichler:2014cwa}. In AdS space, this is not true anymore, but still, each polynomial shift symmetry of \emph{level} $k$ can be kept separately if the free field $\phi$ has a particular mass depending on $k$ \cite{Bonifacio:2018zex, Hinterbichler:2022vcc}.
More precisely, in AdS$_{d+1}$ a free field $\phi$ \footnote{The interactions in DGKT are $1/N$ suppressed, and so the fields are free in the large flux $N \rightarrow \infty$ limit \cite{Conlon:2021cjk}.}
has a symmetry
\begin{aleq}\label{polyshift}
\phi \rightarrow \phi + c_{\mu_1 \dots \mu_k} X^{\mu_1}\dots X^{\mu_k}\vert_{AdS},
\end{aleq}
where $X^\mu$ are coordinates on an embedding $(d+2)$-dimensional flat spacetime, if the mass of the field equals
\begin{aleq}
m_{\phi}^2 = \dfrac{k(k+d)}{R_{AdS}^2},
\end{aleq}
with $R_{AdS}$ the radius of AdS. In a dual CFT, such masses correspond to integer dimensions
\begin{aleq}
\Delta_{+} = k + d, \quad \text{or} \quad \Delta_{-} = -k.
\end{aleq}
We conclude that there are indeed polynomial shift symmetries for the moduli and axions in DGKT related to the integer dimensions:
\begin{aleq}
    & \Delta_{\phi} = \Delta_{+}  = k+3, \quad k = 3, 7,\\
    & \Delta_{a} = \Delta_{+}  = k+3, \quad k = 2, 8.
\end{aleq}
A microscopic explanation of these shift symmetries, possibly related to the domain walls discussed in the next section or the discrete symmetries of \cite{Buratti:2020kda} in the large $N$ limit, would be very interesting.

To check what these symmetries look like on the boundary of AdS, we consider Poincar\'e coordinates
\begin{aleq}
ds_{AdS}^2 = \dfrac{R_{AdS}^2}{z^2} \left( -dx_0^2 + \sum_{i=1}^{d-1} dx_i^2+ dz^2 \right),
\end{aleq}
the boundary being at $z = \epsilon$, $\epsilon \rightarrow 0$. Then the polynomial shifts on the boundary reduce to
\begin{aleq}\label{bdyshift}
    c_{\mu_1 \dots \mu_k} X^{\mu_1}\dots X^{\mu_k}\vert_{AdS} \xrightarrow{z \rightarrow 0}  \left(\dfrac{R_{AdS}}{z}\right)^k c_{i_1 \dots i_k}  X^{i_1} \dots X^{i_k} \vert_{boundary},
\end{aleq}
with $X_i$ coordinates of the flat embedding restricted to the boundary.
Considering the near-boundary behaviour of the scalar field
\begin{aleq}
    \phi \sim \phi_0 z^{\Delta_{-}} + \phi_1 z^{\Delta_{+}} = \phi_0 z^{-k} + \phi_1 z^{k+d},
\end{aleq}
we see that the shift acts only on $\phi_0$ as
\begin{aleq}
    \phi_0 \rightarrow \phi_0 + R_{AdS}^k c_{i_1 \dots i_k}  X^{i_1} \dots X^{i_k} \vert_{boundary}.
\end{aleq}
In standard quantization $\phi_0$ is fixed, and this means that the shift symmetry is broken on the boundary \cite{Blauvelt:2022wwa}.
This should be investigated more precisely when more is known about an explicit DGKT CFT dual.

\section{Large $N$ scalings from D-brane domain walls}
Parametric control, and possibly scale separation, can be obtained in AdS flux vacua in the $N \rightarrow \infty$ limit, where $N$ is proportional to some unbounded flux. This flux, and also the bounded fluxes needed for moduli stabilization, may be interchanged for brane domain walls interpolating between the AdS vacuum and flat spacetime, as in \cite{Kounnas:2007dd}. Interestingly, an AdS$\times (\text{torus})$ geometry may be found in the near-horizon region of this entire brane-system and the dilaton approaches a finite constant when approaching the horizon \cite{Kounnas:2007dd}.

In this section, we focus on the branes dual to the unbounded fluxes and 
explain how one can derive the scalings of the central charge, string coupling, and internal volume for most AdS flux vacua by just looking at the near-horizon geometry of these branes. We first do this for a single stack of domain walls in general flux vacua, and afterwards apply it to some concrete examples including the DGKT and scale-separated AdS$_3$ vacua.
\subsection{General flux vacua}\label{generalflux}
If there is one unbounded flux, $F_{8-p} \sim N$, we can
change this for $N$ D$p$-branes, wrapped on $(p+2-D)$-cycles, in a 10d \textcolor{black}{flat spacetime consisting of $D$ non-compact directions, the other directions forming a toroidal geometry}. These branes can be considered as domain walls in the $D$-dimensional spacetime, with coordinates ($t, x_1, x_2, \dots x_{D-2},x$), and where $x$ will be the direction transverse to the branes. We denote the wrapped internal coordinates by ($y_1, y_2, \dots y_{p+2-D}$), and the unwrapped ones by ($z_1, z_2, \dots z_{8-p}$), as shown in Table \ref{general}.
\begin{table}[h!]
        \begin{center}
            \begin{tabular}{ | l | l | l |l |l |l |l |l |l |l |l |l |}
            \hline
             & $t$ & $x^1$ & $\dots$ & $x^{D-2}$ & $x$ & $y_1$ & $\dots$ & $y_{p+2-D}$ & $z_1$ & $\dots$ & $z_{8-p}$ \\
            \hline
            \textbf{N D$p$} & $\otimes$ & $\otimes$ & $\otimes$ & $\otimes$ &  & $\otimes$  & $\otimes$  & $\otimes$ & & & \\
            \hline
        \end{tabular}
    \caption{$N$ D$p$-brane domain walls in $D$ dimensions}
    \label{general}
    \end{center}
\end{table}

This system is described by \cite{Aharony:1999ti}
\begin{aleq}
    ds_{10}^2 = f_p^{-\frac{1}{2}}\left[ -dt^2 + dx_1^2 + \dots + dx_{D-2}^2\right] +  f_p^{\frac{1}{2}}dx^2 +  f_p^{-\frac{1}{2}}dy_i dy^i + f_p^{\frac{1}{2}}dz_j dz^j,
\end{aleq}
and the string coupling is
\begin{aleq}
    g_s = f_p ^{\frac{3-p}{4}},
\end{aleq}
with 
\begin{aleq}
    f_p \sim 1 + \dfrac{N}{r^{7-p}}, \quad r^2 = \sum_i y_i^2 + \sum_j z_j^2,
\end{aleq}
a \textcolor{black}{harmonic function in the transverse coordinates} that scales like $f_p \sim N$ in the near-horizon region.

Letting $N \rightarrow \infty$, the near-horizon geometry will be of the following schematic form,
\begin{aleq}
    ds_{NH}^2 = \alpha' \left[ N^{\frac{1}{2}} ds_{X_D}^2 + N^{-\frac{1}{2}}dy_i dy^i + N^{\frac{1}{2}}dz_j dz^j \right],
\end{aleq}
where $X_D$ is the $D$-dimensional spacetime and
\begin{aleq}\label{SC1}
    \boxed{g_s \sim N^{\frac{3-p}{4}}}.
\end{aleq}
We see that there is a separation of scales between the wrapped internal dimensions and the scale of $X_D$, and we can read off the internal volume in string units:
\begin{aleq}\label{SC2}
    \boxed{vol_S \sim (N^{-\frac{1}{2}})^{\frac{p+2-D}{2}}\cdot (N^{\frac{1}{2}})^{\frac{8-p}{2}} \cdot l_s^{10-D} = N^{\frac{D+6-2p}{4}}\cdot l_s^{10-D}}.
\end{aleq}
From this, we deduce the D-dimensional Planck length
\begin{aleq}
l_{p,D} = (g_s^2 vol_S^{-1})^{1/(D-2)} l_s = N^{-\frac{D}{4\,\left(D-2\right)}} l_s.
\end{aleq}
Hence, the metric in Planck units is
\begin{aleq}
ds_{NH}^2 = N^{\frac{1}{D-2}}\left[N ds_{X_D}^2 + dy_i dy^i + Ndz_j dz^j\right]l_{p,D}^2.
\end{aleq}
Then, the AdS radius is
\begin{aleq}
R_{AdS} \sim N^{\frac{D-1}{2\,\left(D-2\right)}} l_{p,D},
\end{aleq}
and so the central charge will be
\begin{aleq}\label{SC3}
\boxed{c \sim (R_{AdS}/l_{p,D})^{D-2}\sim N^{\frac{D-1}{2}}}.
\end{aleq}
Repeating the same calculation for domain walls from NS5-branes, we find
\begin{aleq}
    g_s \sim N^{\frac{1}{2}}, \quad vol_S \sim N^{\frac{3}{2}}, \quad c \sim N^{\frac{D-1}{2}},
\end{aleq}
or for F1-strings (only for $D=3$),
\begin{aleq}
    g_s \sim N^{-\frac{1}{2}}, \quad vol_S \sim  N^{\frac{D-3}{2}}, \quad c \sim N^{\frac{D-1}{2}}.
\end{aleq}
In the examples that we discuss below, there will often be multiple stacks like in Table \ref{general} of D-branes dual to unbounded flux. If these stacks are orthogonally-intersecting, they must satisfy certain composition rules, summarised in \cite{Caviezel:2008ik}. For the D$p$/D$q$-brane intersections that we consider, each pair of these branes must have ($0$ mod $4$) relative transverse coordinates. These are coordinates that are orthogonal to one of the two branes, but not both. Then, we can assign a function $f_{p_i}$ to each stack of D$_{p_i}$-branes, that only depends on the overall transverse directions. For each of the coordinate directions $\zeta = t, x, y$ or $ z$, we then multiply the appropriate powers of these functions as follows,
\begin{aleq}\label{zeta}
    \prod_i f_{p_i}^{1/2} \cdot \prod_j f_{q_j}^{-1/2} \cdot (d\zeta)^2,
\end{aleq}
where the $p_i$-branes are orthogonal, and the $q_j$-branes are parallel to the $\zeta$-direction. With these harmonic superposition rules \cite{Tseytlin:1996bh}, we obtain the supergravity description of the brane systems. The string coupling will be given by
\begin{aleq}
    g_s = \prod_i f_{p_i}^{\frac{3-p_i}{4}},
\end{aleq}
where the multiplication is over all branes involved,
and we will deduce the central charge and internal volumes from the near-horizon geometries of \eqref{zeta}.

\subsection{AdS$_4$ vacua}\label{4ddomainwalls}
\subsubsection{Massive DGKT}
The DGKT \cite{DeWolfe:2005uu} potential is given by
\begin{equation}\label{DGKT}
    V = \dfrac{1}{s^3}\left[
\dfrac{A_{F_4}}{us}+ \dfrac{A_{F_0} u^3}{s}+\dfrac{A_{H_3} s}{u^3}-A_{O6}\right],
\end{equation}
where $u^3 = vol_S$ and $s=e^{-D} = e^{-\phi}\sqrt{vol_S}$, and the $A$'s are coefficients depending on the values of the fluxes or the orientifold charge, with $A_{O6}^2 = 16 A_{F_0} A_{H_3}$.
The K\"ahler potential equals
\begin{aleq}
    K = -3 \log u - 4 \log s.
\end{aleq}
The conformal dimensions of the dual fields are
\begin{aleq}
    \Delta_1 = 6, \quad \Delta_2 = 10.
\end{aleq}
Imposing the scaling $A_{F_4}\sim N^2$, we find that, at the minimum of the potential
\begin{aleq}
    V \sim N^{-\frac{9}{2}}, \quad s \sim N^{\frac{3}{2}}, \quad u \sim N^{\frac{1}{2}},
\end{aleq}
which means that
\begin{aleq}\label{dgktscalings}
    c\sim N^{9/2}, \quad g_s \sim N^{-3/4}, \quad vol_S \sim N^{3/2}.
\end{aleq}
If the internal manifold is a toroidal orientifold $T^6/\mathbb{Z}_3^2$, $F_{4,i} \sim N_i $ ($ i =1,2,3$) flux is turned on along each of the 3 sub-2-tori. We can interchange these 3 fluxes for 3 stacks domain walls consisting of respectively $N_i$ D4-branes wrapped on the different 2-cycles, as shown in Table \ref{dgkt}.
\begin{table}[h!]
        \begin{center}
            \begin{tabular}{ | l | l | l |l |l |l |l |l |l |l |l |}
            \hline
             & $t$ & $x^1$ & $x^2$ & $x$ & $y_1$ & $y_2$ & $y_3$ & $y_4$ & $y_5$ & $y_6$ \\
            \hline
            \textbf{$N_1$ D4} & $\otimes$ & $\otimes$ & $\otimes$  & &$\otimes$ &  $\otimes$  && && \\
            \hline
            \textbf{$N_2$ D4} & $\otimes$ & $\otimes$ & $\otimes$  & &&    &$\otimes$&$\otimes$ && \\
            \hline
           \textbf{$N_3$ D4} & $\otimes$ & $\otimes$ & $\otimes$  & &&    && &$\otimes$& $\otimes$\\
            \hline
        \end{tabular}
    \caption{D4-brane domain walls in DGKT}
    \label{dgkt}
    \end{center}
\end{table}

As explained in \cite{Kounnas:2007dd}, we can also change the other bounded $H_3$- and $F_0$-fluxes for NS5-branes wrapped on 3-cycles and D8-branes on 6-cycles respectively. The resulting near-horizon geometry of this brane set-up will be AdS$_4 \times T_6$. In the following, we will not explicitly take the presence of these sub-leading branes, dual to bounded fluxes, into account, but just assume they are there to make  the near-horizon geometry of the right form and look only at the large $N$ dependence of this geometry.

As $N_i \sim N \rightarrow \infty$, using the harmonic superposition rules, we find
\begin{aleq}
    ds_{NH}^2 &= \alpha' \big[ N_1^{\frac{1}{2}}N_2^{\frac{1}{2}}N_3^{\frac{1}{2}}ds_{X_4}^2 + N_1^{-\frac{1}{2}}N_2^{\frac{1}{2}}N_3^{\frac{1}{2}}(dy_1^2+dy_2^2)\\ & + N_1^{\frac{1}{2}}N_2^{-\frac{1}{2}}N_3^{\frac{1}{2}}(dy_3^2+dy_4^2) + N_1^{\frac{1}{2}}N_2^{\frac{1}{2}}N_3^{-\frac{1}{2}}(dy_5^2+dy_6^2) \big],
\end{aleq}
with
\begin{aleq}
    g_s \sim N_1^{-\frac{1}{4}}N_2^{-\frac{1}{4}}N_3^{-\frac{1}{4}} \sim N^{-\frac{3}{4}}.
\end{aleq}
The internal volume can be read off,
\begin{aleq}
    vol_S \sim N_1^{\frac{1}{2}}N_2^{\frac{1}{2}}N_3^{\frac{1}{2}} \sim N^{\frac{3}{2}},
\end{aleq}
along with the degree of scale separation,
\begin{aleq}
    \dfrac{L_{KK}^2}{L_{H}^2} \sim \text{max}_i (N_i^{-1}) \sim N^{-1}.
\end{aleq}
The 4d Planck scale is given by
\begin{aleq}
    l_{p,4} = \dfrac{g_s}{\sqrt{vol_S}} l_s \sim N_1^{-\frac{1}{2}}N_2^{-\frac{1}{2}}N_3^{-\frac{1}{2}} l_s,
\end{aleq}
and so it follows that the central charge is given by
\begin{aleq}
    c \sim \dfrac{R_{X_4}^2}{l_{p,4}^2} \sim N_1^{\frac{3}{2}}N_2^{\frac{3}{2}}N_3^{\frac{3}{2}} \sim N^{\frac{9}{2}}.
\end{aleq}
We obtain the same scalings as the ones \eqref{dgktscalings} from the scalar potential.

\subsubsection{Massless DGKT}
We can proceed similarly for DGKT vacua without Romans mass. These are obtained by performing two T-dualities on massive DGKT and then re-scaling fluxes to get a solution under control \cite{Cribiori:2021djm, Caviezel:2008ik}. The internal manifold will now have curvature becoming a more general SU(3) structure manifold. The flux distribution is anisotropic, with only unbounded flux on 2 out of 3 2-tori in the toroidal case. That is why we now write explicitly $vol_s = u_1 u_2 u_3$. There is unbounded $F_6$-flux, both bounded $F_2$-flux ($F_{2,1}$) and unbounded $F_2$-flux ($F_{2,2}, F_{2,3}$), a curvature $R$ contribution and again there are O6-planes. The scalar potential is given by
\begin{aleq}
    V = \dfrac{1}{s^3} \left[ A_{F_6} \dfrac{1}{s u_1 u_2 u_3}+ A_{F_{2,1}} \dfrac{u_2 u_3}{s u_1} + A_{F_{2,2}} \dfrac{u_1 u_3}{s u_2}+ A_{F_{2,3}} \dfrac{u_1 u_2}{s u_3}+A_R \dfrac{s u_1}{u_2 u_3}-A_{O6}\right],
\end{aleq}
with K\"ahler potential 
\begin{aleq}
    K = -\log u_1 u_2 u_3 - 4 \log s.
\end{aleq}
Conformal dimensions do not change under T-duality:
\begin{aleq}
    \Delta_{1,2,3} = 6, \quad \Delta_4 = 10.
\end{aleq}
Letting $F_6\sim N$ , $F_{2,2} \sim M_1$ and $F_{2,3} \sim M_2$, it follows that
\begin{aleq}\label{TTscalings}
    c \sim N^{\frac{3}{2}} M_1^{\frac{3}{2}} M_2^{\frac{3}{2}}, \quad g_s \sim N^{\frac{1}{4}} M_1^{-\frac{3}{4}} M_2^{-\frac{3}{4}}, \quad vol_S \sim N^{\frac{3}{2}} M_1^{-\frac{1}{2}} M_2^{-\frac{1}{2}}.
\end{aleq}
For $N$ large enough compared to $M_1$ and $M_2$, the coupling will be strong and this solution can be uplifted to M-theory. \textcolor{black}{The first dimension-6 operator can then be interpreted as being dual to the (7d) internal volume.} 

Now we change the $F_6$-flux for D2-branes, and the unbounded $F_2$-fluxes for D6-branes wrapped on 4-cycles, as in Table \ref{TTdgkt}.
\begin{table}[h!]
\begin{center}
    \begin{tabular}{ | l | l | l |l |l |l |l |l |l |l |l |}
    \hline
     & $x^0$ & $x^1$ & $x^2$ & $x$ & $y_1$ & $y_2$ & $y_3$ & $y_4$ & $y_5$ & $y_6$  \\
    \hline
    \textbf{$N$ D2} & $\otimes$ & $\otimes$ & $\otimes$ & & & &&   & & \\
    \hline
        \textbf{$M_1$ D6} & $\otimes$ & $\otimes$ & $\otimes$ & &$\otimes$ & $\otimes$&$\otimes$ & $\otimes$&&    \\
    \hline
        \textbf{$M_2$ D6} & $\otimes$ & $\otimes$ & $\otimes$ & &$\otimes$ & $\otimes$&& &$\otimes$ & $\otimes$  \\
          \hline
\end{tabular}
\caption{D2- and D6-brane domain walls in massless DGKT}
\label{TTdgkt}
\end{center}
\end{table}

The following near-horizon geometry can be read off:
\begin{aleq}
    ds_{NH}^2 &= \alpha' \big[ N^{\frac{1}{2}}M_1^{\frac{1}{2}}M_2^{\frac{1}{2}} ds_{X_4}^2 + N^{\frac{1}{2}}M_1^{-\frac{1}{2}}M_2^{-\frac{1}{2}} (dy_{1}^2+dy_2^2)\\& +  N^{\frac{1}{2}}M_1^{-\frac{1}{2}}M_2^{\frac{1}{2}}(dy_{3}^2+dy_4^2) +  N^{\frac{1}{2}}M_1^{\frac{1}{2}}M_2^{-\frac{1}{2}} (dy_{5}^2+dy_6^2)\big],\\
\end{aleq}
with
\begin{aleq}
 g_s  \sim N^{\frac{1}{4}} M_1^{-\frac{3}{4}} M_2^{-\frac{3}{4}},\quad \dfrac{L_{KK}^2}{L_{H}^2} \sim \text{max}_i (M_i^{-1}),
\end{aleq}
fully in accordance with the scalings from the scalar potential \eqref{TTscalings}.
\subsubsection{Other parametric AdS$_4$ vacua}
These large $N$ scalings from domain walls do not only work out for scale-separated vacua as in the examples above, but also for more general 4d examples with parametric control without scale separation. A simple example is obtained by imposing isotropy $u_1=u_2=u_3$ in the potential of the last section, so that
\begin{aleq}
    V = \dfrac{1}{s^3} \left[ A_{F_6} \dfrac{1}{s u^3}+ A_{F_{2}} \dfrac{u}{s}+A_R \dfrac{s}{u}-A_{O6}\right],
\end{aleq}
and then the only unbounded flux will be $F_6 \sim N$. The conformal dimensions are now
\begin{aleq}
    \Delta_1 = 6, \quad \Delta_2 = 11/3.
\end{aleq}
The scalings are
\begin{aleq}
    c \sim N^{\frac{3}{2}}, \quad g_s \sim N^{\frac{1}{4}}, \quad vol_S \sim N^{\frac{3}{2}},
\end{aleq}
which coincide with the scalings \eqref{SC1}, \eqref{SC2}, \eqref{SC3} for one stack of $N$ D2-branes in 4d. There is clearly no scale separation, as can be seen from the near-horizon geometry as well,
\begin{aleq}
    ds_{NH}^2 = \alpha'[ N^{\frac{1}{2}}ds_{X_4}^2 +  N^{\frac{1}{2}}(dy_1^2+dy_2^2+dy_3^2+dy_4^2+dy_5^2+dy_6^2)].
\end{aleq}
\textcolor{black}{Reference \cite{Grimm:2019ixq} classifies 2-moduli scalar potentials for asymptotic 4d $\mathcal{N}=1$ flux vacua. For the AdS vacua they obtain near infinite distance singularities, we observe that}
\begin{aleq}
    \Delta_1 =6, \quad \Delta_2 \ \text{is rational}.
\end{aleq}
Moreover, the central charge is always given by
\begin{aleq}
    c \sim N^{\frac{3k}{2}},
\end{aleq}
with $N$ an unbounded flux and $k$ some positive integer. With \eqref{SC3}, this suggests that these vacua can be obtained in the near-horizon limit of $k$ stacks of D-brane domain walls. \textcolor{black}{The string coupling and internal volume scale accordingly}. We will report on this in more detail in the future.
\subsection{AdS$_3$ vacua}\label{3ddomainwalls}
\subsubsection{Scale-separated AdS$_3$ in massive IIA}
The vacua of \cite{Farakos:2020phe} are scale-separated AdS$_3$ vacua, from compactification of massive IIA string theory on a G2 holonomy manifold. There are $F_4$-, $H_3$- and $F_0$- fluxes, together with O6-planes, and the scalar potential is:
\begin{equation}\label{V3dIIA}
   V = \dfrac{A_{F_4}}{u^{\frac{1}{2}}s^3}+\dfrac{A_{F_0}u^{\frac{7}{2}}}{s^3}+ \dfrac{A_{H_3}}{u^3 s^2}-\dfrac{A_{O6}u^{\frac{1}{4}}}{s^{\frac{5}{2}}}.
\end{equation}
with  $A_{O6}^2 = 12\cdot A_{f_0}\cdot A_{h_0}$, and where $u^{7/2} = vol_S$ and $s = e^{-2 \phi} vol_S$ \footnote{To make contact with the notation in \cite{Farakos:2020phe}: $u=e^{\frac{\phi}{2}+2\beta v}$, $s= e^{-\frac{\phi}{4}+7\beta v}$.}. The conformal dimensions are irrational
\begin{aleq}
    \Delta = 1 + \sqrt{\dfrac{191\pm 8 \sqrt{277}}{7}}.
\end{aleq}
With $F_4 \sim N$, we find from \eqref{V3dIIA} that
\begin{aleq}\label{3dscalings}
    c \sim N^4, \quad g_s \sim N^{-3/4}, \quad vol_S \sim N^{7/4}.
\end{aleq}
There are such $F_4$-fluxes on seven different 4-cycles, and interchanging them for D4-branes wrapped on the dual 3-cycles, we find the system of Table \ref{FTV}.
\begin{table}[h!]
        \begin{center}
            \begin{tabular}{ | l | l | l |l |l |l |l |l |l |l |l |}
            \hline
             & $t$ & $x^1$ & $x$ & $y_1$ & $y_2$ & $y_3$ & $y_4$ & $y_5$ & $y_6$ & $y_7$\\
            \hline
            \textbf{$N_1$ D4} & $\otimes$ & $\otimes$ &  &$\otimes$ &$\otimes$ &  && && $\otimes$ \\
            \hline
            \textbf{$N_2$ D4} & $\otimes$ & $\otimes$ & & & & $\otimes$& $\otimes$&    & &$\otimes$\\
            \hline
            \textbf{$N_3$ D4} & $\otimes$ & $\otimes$ & & & & &   & $\otimes$& $\otimes$ &$\otimes$\\
            \hline
        \textbf{$N_4$ D4} & $\otimes$ & $\otimes$ & & $\otimes$& &$\otimes$& &   &$\otimes$ & \\
            \hline
        \textbf{$N_5$ D4} & $\otimes$ & $\otimes$ & & & $\otimes$& $\otimes$ &  & $\otimes$& &\\
            \hline
        \textbf{$N_6$ D4} & $\otimes$ & $\otimes$ & & $\otimes$& & &  $\otimes$ & $\otimes$& &\\
            \hline
        \textbf{$N_7$ D4} & $\otimes$ & $\otimes$ & & & $\otimes$&  &  $\otimes$ & & $\otimes$ &\\
            \hline
        \end{tabular}
        \caption{D4-brane domain walls for scale-separated AdS$_3$}
        \label{FTV}
    \end{center}
\end{table}

There should be further domain walls from NS5-branes and D8-branes, dual to the bounded $H_3$ - and $F_0$-fluxes. If there is indeed an AdS$_3 \times T_7$ geometry in the near-horizon limit of this set of domain walls, the schematic flux dependence should be like
\begin{aleq}
    ds_{NH}^2 &= \alpha' \big[ (N_1 \cdot \dots \cdot N_7)^{\frac{1}{2}}ds_{X_3}^2 + (N_1 N_4 N_6)^{-\frac{1}{2}}(N_2 N_3 N_5 N_7)^{\frac{1}{2}} dy_1^2 + \dots \big]\\
    &= \alpha' \big[ N^{\frac{7}{2}}ds_{X_3}^2 + N^{\frac{1}{2}} (dy_1^2+dy_2^2+dy_3^2+dy_4^2+dy_5^2+dy_6^2+dy_7^2)  \big],\\
\end{aleq}
with $N_i=N, i = 1, \dots 7$. The string coupling \eqref{SC1} would be
\begin{aleq}
    g_s = (N_1  \cdot \dotsc \cdot   N_7)^{-\frac{1}{4}} = N^{-\frac{7}{4}},
\end{aleq}
and the internal volume is
\begin{aleq}
    vol_S = (N_1  \cdot \dotsc \cdot   N_7)^{\frac{1}{4}} = N^{\frac{7}{4}}.
\end{aleq}
Finally, the central charge would be
\begin{aleq}
    c = N_1  \cdot \dotsc \cdot   N_7 = N^7.
\end{aleq}
Remarkably, the scalings of the central charge and the string coupling, derived from the brane system do not match with \eqref{3dscalings}.
Given that each of the seven internal lengths should scale in the same way with $N$ \cite{Farakos:2020phe}, and taking into account that each pair of stacks of D4-branes should have $0$ or $4$ relative transverse directions \cite{Caviezel:2008ik}, the set-up of Table \ref{FTV} is the only possible set-up consisting of orthogonally-intersecting D4-branes.
\subsubsection{Parametric AdS$_3$ in IIB}
To stress that this failure to obtain the scalings from the near-horizon geometry of a simple system of orthogonally-intersecting branes is not generic to asymptotic vacua in 3d, we mention another example: AdS$_3$ vacua from compactification of IIB string theory on G2-structure manifolds (so with curvature $R$) \cite{Emelin:2021gzx} with $F_7$-, $F_3$-fluxes and O5-planes and possibly D5-branes. These do not allow scale separation, but the moduli can take parametric values. The scalar potential equals
\begin{equation}\label{V3dIIB}
   V = \dfrac{A_{F_7}}{u^{\frac{7}{2}}s^3}+\dfrac{A_{F_3}u^{\frac{1}{2}}}{s^3}+ \dfrac{A_{R}}{u s^2}-\dfrac{A_{O5/D5}}{u^{\frac{1}{4}}s^{\frac{5}{2}}},
\end{equation}
with $A_{O5/D5}^2 = 16 A_{F3} \cdot A_R/3$.
The conformal dimensions are now integer and rational 
\begin{aleq}
    \Delta_1 = 4, \quad \Delta_2 = \dfrac{20}{7}.
\end{aleq}
Choosing $F_7 \sim N$, the scalings
\begin{aleq}
    c\sim N, \quad g_s \sim N^{1/2}, \quad vol_S \sim N^{7/4}.
\end{aleq}
from the scalar potential \eqref{V3dIIB}. These coincide with the scalings \eqref{SC1}, \eqref{SC2}, \eqref{SC3} for $N$ D1-domain walls in 3d, see Table \ref{Bads3}.
\begin{table}[h!]
        \begin{center}
            \begin{tabular}{ | l | l | l |l |l |l |l |l |l |l |l |}
            \hline
             & $t$ & $x^1$ & $x$ & $y_1$ & $y_2$ & $y_3$ & $y_4$ & $y_5$ & $y_6$ & $y_7$\\
            \hline
            \textbf{$N$ D1} & $\otimes$ & $\otimes$ &  && &  && &&  \\
            \hline
        \end{tabular}
    \caption{D1-brane domain walls for parametric AdS$_3$}
    \label{Bads3}
    \end{center}
\end{table}
\section{Conclusions}
The holographic duals of the DGKT vacua are remarkable, in the sense that their spectrum consists of integer dimensions and that the scaling of the central charge $c\sim N^{9/2}$ is unseen in any well-understood holographic set-up. We have observed, however, that the DGKT scalings are consistent with the large $N$ near-horizon limit of three intersecting stacks of $N$ D4-branes wrapped on 2-cycles. Similarly, the scalings in massless DGKT, obtained after twice T-dualizing the original DGKT vacua \cite{Cribiori:2021djm}, are consistent with the near-horizon geometry of one stack of D2-branes and two stacks of D6-branes wrapped on 4-cycles.

More generally, AdS vacua in $(d+1)$ dimensions with minimal supersymmetry and an unbounded flux proportional to $N$, and where the conformal dimensions take specific values
\begin{aleq}\label{dims}
    \Delta_1 = 2d, \quad \Delta_2 \quad \text{is \ rational}
\end{aleq}
seem to agree with a near-horizon geometry of $\mathcal{O}(N)$ D-brane domain walls as $N \rightarrow \infty$, whereas this does not seem the case for vacua with irrational dimensions. The integer dimension $\Delta_1 = 2d$ signals the presence of a level $d$ polynomial shift symmetry for the corresponding field.

The scale-separated AdS$_3$ vacua are unlike their 4-dimensional DGKT counterparts, despite being built from similar ingredients, with irrational dimensions and scalings that cannot be obtained from orthogonal D4-domain walls. Moreover, a massless version ($F_0 \neq 0$) does not exist here \cite{VanHemelryck:2022ynr}, preventing an M-theory uplift \cite{Aharony:2010af}. Future research will clarify the origin of the failed scalings for the AdS$_3$ vacua and whether they imply any inconsistency.

\subsection*{Acknowledgements}
I am grateful to Thomas Van Riet, Vincent Van Hemelryck, Miguel Montero, Irene Valenzuela, George Tringas, and especially Joe Conlon for useful discussions and comments on the draft. I am supported by the Clarendon Scholarship in partnership with the Scatcherd European Scholarship, Saven European Scholarship, and the Hertford College Peter Howard Scholarship. 

\small{\bibliography{refs}}
\bibliographystyle{utphys}
\end{document}